\documentclass{article}
\usepackage{arxiv}

\usepackage[utf8]{inputenc} 
\usepackage[T1]{fontenc}    
\usepackage[hidelinks]{hyperref}   
\usepackage{url}            
\usepackage{booktabs}       
\usepackage{amsfonts}       
\usepackage{nicefrac}       
\usepackage{microtype}      
\usepackage{lipsum}		
\usepackage{graphicx}
\usepackage{doi}
\usepackage{amsmath}
\usepackage{lipsum} 
\usepackage{graphicx} 
\usepackage{setspace}
\usepackage{subfig}
\usepackage{environ} 
\usepackage{bbm}
\usepackage[labelformat=empty]{caption}
\usepackage{float}
\usepackage{enumerate}
 
\NewEnviron{HUGE}{%
    \scalebox{1.2}{$\BODY$} 
} 

\DeclareMathOperator*{\argmin}{argmin}

\DeclareMathOperator*{\prob}{\mathbb{P}}

\title{Deep Learning Methods for Protein Family Classification on PDB Sequencing Data}

\author{{\hspace{1mm}Aaron Wang} \\
	Department of Computer Science\\
	Brown University\\
	Providence, RI 02912 \\
	\texttt{aaronjwang@brown.edu} \\
}

\date{}

\renewcommand{\headeright}{Final Paper}
\renewcommand{\undertitle}{Final Paper}
\renewcommand{\shorttitle}{Deep Learning on PDB Sequencing Data}

\hypersetup{
pdftitle={Deep Learning Methods for Protein Family Classifcation on PDB Sequencing Data},
pdfsubject={q-bio.NC, q-bio.QM},
pdfauthor={Aaron ~Wang},
pdfkeywords={First keyword, Second keyword, More},
}
\begin{document}
\maketitle

\begin{abstract}
Composed of amino acid chains that influence how they fold and thus dictating their function and features, proteins are a class of macromolecules that play a central role in major biological processes and are required for the structure, function, and regulation of the body’s tissues. Understanding protein functions is vital to the development of therapeutics and precision medicine, and hence the ability to classify proteins and their functions based on measurable features is crucial; indeed, the automatic inference of a protein's properties from its sequence of amino acids, known as its primary structure, remains an important open problem within the field of bioinformatics, especially given the recent advancements in sequencing technologies and the extensive number of known but uncategorized proteins with unknown properties. In this work, we demonstrate and compare the performance of several deep learning frameworks, including novel bi-directional LSTM and convolutional models, on widely-available sequencing data from the Protein Data Bank (PDB) of the Research Collaboratory for Structural Bioinformatics (RCSB), as well as benchmark this performance against classical machine learning approaches, including k-nearest neighbors and multinomial regression classifiers, trained on experimental data. Our results show that our deep learning models deliver superior performance to classical machine learning methods, with the convolutional architecture providing the most impressive inference performance.
\end{abstract}


\section*{Introduction}

Proteins are macromolecules responsible for the function and regulation of the body's tissues and organs. At their most basic level, proteins are composed of chains of amino acids, called polypeptide chains, that themselves are determined by the sequence of nucleotides in a gene. Two proteins that diverged from a common ancestral family are known as homologous and tend to have similar structures, functions, and general behavior; when a new protein is discovered, its functional and structural properties can be suggested based on what group of prior labelled proteins the new proteins is most similar to. This is the goal of protein classification. Classification techniques have been used ubiquitously in statistics, computer science, biology, and the social sciences, and are powerful for classifying samples within data into classes with similar patterns; this branch of machine learning, known as supervised learning, has proven useful for accurate prediction of group labels in protein classification where proteins in the same class likely share similar functionality and structure. The application of classification methods to protein homology modeling is an important task since millions of proteins have been studied but unsuccessfully classified into protein families. For instance, well-curated Swiss Prot dataset contains 560,000 annotated proteins, whereas the sparsely annotated TrEMBL dataset contains over 150 million proteins.

\section*{Approach}
\label{sec:headings}
\subsection*{Deep Learning}
Deep learning is a specific family of models within machine learning that relies on artificial neural networks, an algorithmic system of artificial neurons that processes information and learns complex patterns in data, and is broadly inspired by biological systems in the human brain; indeed, these models have become exceptionally popular in the bioinformatics community in recent years because of their ability to extract conclusive and often complex patterns from large datasets. \cite{bib27} first used feed forward neural network for protein structure prediction, and since then neural networks have been used for a number of applications in the field of genomics, which include predicting the outcome of alternative splicing in exon skipping events ~\cite{bib12}, annotating coding and non-coding genetic variants for identifying pathogenic variants ~\cite{bib13}, and identifying active cis-regulatory regions in the human genome (~\cite{bib14}, ~\cite{bib15}). Many of these methods report improved predictive performance over methods such as linear regression, k-nearest neighbors clustering, and support vector machines. In this work, we consider two types of neural network architectures that differ in how the data is processed within the models: recurrent neural networks (RNN) and convolutional neural networks (CNN).

\subsubsection*{Recurrent Neural Networks}
Recurrent neural networks, also known as RNNs, are a class of neural networks that introduce state variables to store past information, together with the current inputs, to determine outputs. As such, RNNs are are designed for sequential or time-series data, as hidden layers of the RNN can be viewed as memory states that retain information from the sequence previously observed and are updated at each time step. More specifically, RNNs process the input sequentially, such that at each time step, the RNN takes as input the current subsequence of the input and output of the past layer; what results from these operations is a context matrix contains the information of the sequence, which can be used as input to another model, such as a feedforward neural network. This architecture suffers from the vanishing gradient problem, wherein the weights of the model are unable to learn long-term sequential patterns due to the exponential shrinking of the gradient per time step. To remediate this we utilize a variation of the RNN, known as a bidirectional LSTM, that make use of gating procedures that regulate the neural network's ability to model long-term dependencies.

In genomics, RNNs have been used in applications including prediction of single-cell DNA methylation states from CpG coverage \cite{bib16} and prediction of non-coding function de novo from DNA sequences \cite{bib17}.

\subsubsection*{Convolutional Neural Networks}
In a convolutional neural network (CNN), a convolutional filter, or kernel, is scanned across the entire input matrix and computes the local weighted sum of each region. Each convolutional layer scans the input with several of these filters, each producing a scalar value at every position that denotes the "match" between the input and the kernel- that is, each kernel is initialized separately and attempts to learn a distinct spatial feature of the data. After the convolutional layer, a nonlinear activation function is applied, often in conjunction with dropout and batch normalization to prevent overfitting. Furthermore, a pooling layer, which reduces the spatial size of the input with the purpose of denoising the data, is applied. This set of layers is known as a convolutional block, and are repeated in a network architecture to enhance the signal produced by the CNN; the output of the final block can be used as input to a multi-layer perception for downstream classification.

DeepBind \cite{bib23}, Basset \cite{bib24}, and DeepSea \cite{bib25} were the first to apply CNN methods genomic datasets. The authors of DeepBind trained several single-task models for binary classification to predict binding affinities of transcription factors, showing that their deep learning approach performed consistently better than existing state-of-the-art classical machine learning approaches. Similarly, the authors of Basset utilize a CNN to predict DNA accessibility features, again modeling the task as binary classification. The DeepSEA model, the largest model of the three with over 50 million trainable parameters, predicts the binary presence or absence of chromatin features, such as transcription factor binding, DNA accessibility, and histone modification.
In this work, we model the task as a multi-class classification problem, as opposed to binary classification. We transform our sequential data by embedding each token in the sentence and using this transformed matrix as input to the first convolutional layer and passing in the out of the final convolutional block as input to the feed forward neural network for prediction. This technique of convolution across the encoded matrix is also seen in DeepChrome \cite{bib26}, where the model's kernels slide across the one-hot encoded DNA sequences to predict gene expression on histone modification data.

\subsection*{Classical Machine Learning}
Machine learning (ML) is a branch of artificial intelligence (AI) where computer systems are able to automatically learn and improve from patterns and observations in data without being explicitly programmed [12]. Machine learning methods lend themselves well to tasks that require the compilation of a list of sequence elements into a set of classes, and thus machine learning techniques have extensively been applied to areas of genomics and genetics. For example, machine learning methods have been used to identify splice sites \cite{bib18}, enhancers \cite{bib20}, promoters \cite{bib19}, and positioned nucleosomes \cite{bib21}, as well as used similar techniques to identify locations of transcription start sites in genomic sequences \cite{bib22}. The effectiveness of machine learning on these tasks are only enabled by the presence of large, complex datasets, many of which have only recently become available through international collaborative projects, such as the NIH’s 4D Nucleome Initiative, ENCODE, the Roadmap Epigenomics Project, and the 100000 Genomes Project. In this work, we utilize data from the Protein Data Bank to predict protein sequences into a set of protein family classes from experimental data using two well-studied machine learning models, multinomial regression and k-nearest neighbors, as a form of comparison against our deep learning models.

\subsubsection*{Multinomial Regression}
Multinomial regression is a generalization of logistic regression to allow for more than two possible categorical labels. In this section, we provide brief analysis and intuition for the underlying mathematics behind the model.

Let $X = \{x_1, x_2, ..., x_n \}$ in $\mathbb{R}^{n \times d}$ for $d$ features and $Y = \{y_1, y_2, ..., y_n\}$ in $\mathbb{R}^{n \times 1}$ where $y_i \in \{1, 2, ..., k\}$ for $k$ classes. We would like to utilize statistical machine learning for multi-class classification such that $X \xrightarrow{model} Y$, and multinomial regression is an intuitive approach, 
since the generated prediction is the label with the maximum predicted probability, itself coming from the softmax function. The parameters of the model are optimized to maximize the likelihood of the observed data.
\\

\begin{figure}[H]
\center
\includegraphics[scale = 0.325]{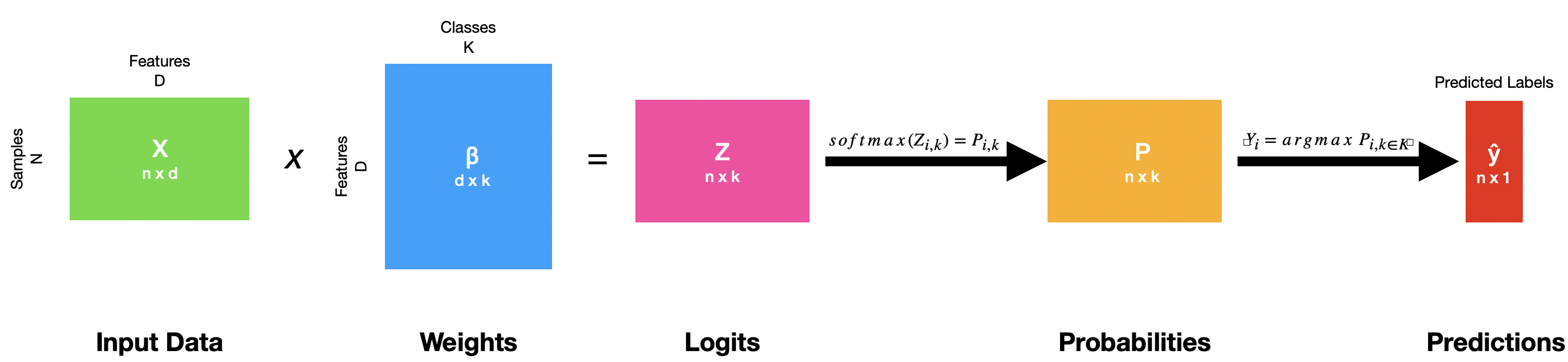}
\caption{Figure 1: Multinomial Regression Overview \protect \footnotemark}
\end{figure}
\footnotetext{the omission of $\alpha$ is done to conserve space for the figure and does not affect the Big O runtime, space complexity, or inherent logic of the model.}

Similar to logistic regression, multinomial regression begins by calculating the logit matrix $Z$ in $\mathbb{R}^{n \times k}$. Note that this is a \textit{matrix} since with $K$ classes, whereas logistic regression calculates a logit vector for $k=2$ and $\prob(Y=0 | X= x) = 1 - \prob(Y = 1 | X=x)$. Outputting the logits directly is not possible since the probabilities outputted by our model before $argmax$ must be non-negative and sum to 1, as defined by the axioms of probability; instead, the model exponentiates each logit to ensure non-negativity and divides by their sum to ensure a summation of 1. This is the intuition behind the softmax function as first defined by Robert Duncan Luce in 1959~\cite{bib8} and is what is used to calculate $\prob(Y_i = k | X = x_i)$. The multinomial regression is still a linear model despite the softmax function being non-linear, as the outputs of multinomial regression are determined by an affine transformation of the feature set.\\

\begin{figure}[H]
	\begin{equation} \label{eqn}
	\prob(Y_i = k | X = x_i) = softmax(Z_{i,k}) = \frac{e^{Z_{i,k}}}{\sum_{k=1} ^{K} e^{Z_{i,k}}}
	\end{equation}
\end{figure}

When $K=2$, $\prob(Y_i = 1 | X = x_i) = \frac{e^{Z_{i,1}}}{e^{Z_{i,1}} + e^{Z_{i,2}}} = \frac{e^{Z_{i,1}}}{e^{Z_{i,1}}(1 + e^{Z_{i,2} - Z_{i,1}})} = \frac{1}{1 + e^{-Z'}}$, which is of the same form as the sigmoid function for logistic regression\footnote{$Z' = Z_{i,1} - Z_{i,2}$. There are an infinite number of combinations of $\beta_1$ and $\beta_2$ that will produce the same probabilities for $Y_1$ and $Y_2$, so $\beta_2$ can be set to $0$ such that $Z' = Z_{i,1}$.}, the only difference being that multinomial is generalized to $K>2$ classes.\\
With regards to defining a loss function to quantify the performance of the model's parameter choices, cross-entropy loss is most often used in literature with multinomial regression since it is derived from MLE~\cite{bib3}, which is straight-forward and only requires some knowledge in information theory; a more interesting derivation of cross-entropy loss comes from KL divergence, a statistical measure of how different two probability distributions are~\cite{bib6}. We can apply this to the true class distribution (denoted $\prob(y|x_i)$) and the predicted class distribution (denoted $\prob(\hat{y}|x_i)$) and attempt to minimize the KL divergence.

\begin{equation} \label{eq:2}
    D_{KL}(\prob(y|x_i) || \prob_{\theta}(\hat{y}|x_i)) \stackrel{\triangle}= \sum_{i=1}^n \prob(y|x_i) \times \log \frac{\prob(y|x_i)}{\prob_\theta(\hat{y}|x_i)}
\end{equation}

After simplification, only $-\sum_{i=1}^n \prob(y|x_i) \times \log \prob_{\theta}(\hat{y}|x_i)$ depends on parameter $\theta$. Thus, $\argmin \limits_{\theta} D_{KL}(\prob(y|x_i) || \prob_{\theta}(\hat{y}|x_i)) = \argmin \limits_{\theta} -\sum_{i=1}^n \prob(y|x_i) \log \prob_{\theta}(\hat{y}|x_i)$. Naturally, this result hints that maximizing likelihood is equivalent to minimizing KL divergence, a notion shown by~\cite{bib6}; that is, maximizing the likelihood of our observed data under our model is equivalent to minimizing the difference between our model's prediction and the true data distribution.

Calculating the root of the derivative of cross-entropy loss with respect to the parameters is \textit{difficult}~\cite{bib9}, so iterative optimization algorithms are deployed to approximate values of the parameters, such as Broyden–Fletcher–Goldfarb–Shanno algorithm, the Newton–Raphson method, and quasi versions of the latter. In this work, we use Newton-CG which applies the linear conjugate gradients to the second-order Taylor-series approximations of $\prob$ around $\theta_i$; it has shown to converge well and in reasonable time complexity as each iteration requires the computation of the Hessian Matrix. One drawback of the Newton-CG method is that saddle points are easier to fall into~\cite{bib5}, but the cross entropy loss function is convex, so this will not be an issue. Our results show that the choice of optimizer does not affect our experiment significantly.\\
Multinomial regression is appropriate for protein classification because the explanatory variables of the data are all ordinal and the response variable is categorical, which is what the model expects as input and outputs, respectively. Furthermore, the proteins were independently sampled from one another and all redundant samples are removed from the data. The features themselves do not show perfect multicollinearity, which is also an assumption of the model that is met. In addition to these assumptions of the model that are met, the model specificically does not make several assumptions which also lends itself well to the protein classification task: Multinomial regression does not require homoscedasticity or that the residuals be normally distributed, an assumption that OLS regression requires~\cite{bib4}.

\subsubsection*{K-Nearest Neighbors}
The K-Nearest Neighbor (KNN) algorithm operates on the basis that similar samples exist in close proximity to one another in Euclidean space; that is, a data point is classified as the label that is most frequent among its k nearest neighbors, where k is a hyperparameter tuned by the researcher. More specifically, let $C$ be the set of $k$ samples closest to point $X_i$:

\begin{equation}
\prob(Y_i = y | X = x) = \frac{1}{K} \sum_{X_j \in C} \mathbbm{1} (Y_j = y)
\end{equation}

The $k$ "closest" samples are determined by how we define distance, though common metrics include Euclidean, Manhattan, Chebyshev, and Hamming distance. The choice of distance metric depends on the data, such as the frequency of outliers you expect, the dispersion of the data, etc.; this work utilizes Euclidean distance since the dimensionality of the data is not high, in which we may look to another metric like Manhattan distance, and Euclidean distance is by far the most popular~\cite{bib10}. The most optimal choice of k depends on the data and the task, although smaller values of k increase the effect of the variance of the samples on the classification. In this work, we tested values of k up to 30 and found that k=3 works best for this task.

\begin{figure}[H]
\center
\includegraphics[scale = 0.4]{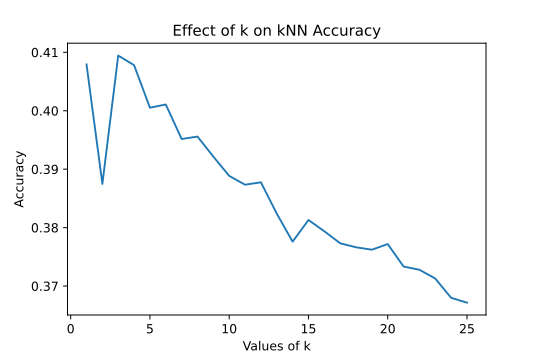}
\caption{Figure 2: Parameter Tuning for k}
\end{figure}

Note that no distinct training is necessary, since the label of a sample is generated by Euclidean distance between the sample and every point. The precomputing of these distances, as well as the use of a k-d tree data structure, are options to increase the efficiency of this algorithm. The inferencing pseudocode for this algorithm is outlined below.

\begin{figure}[H]
Load the data $(x_1,y_1), (x_2,y_2), ..., (x_n, y_n)$, initialize the value $K = k$, and define a distance metric $d$. To determine the label $y_l$ of a new sample $x_l$:

\begin{enumerate}[Step 1:]
\item Compute and store $d(x_l, x_j) \forall x_j \in X$ in $L$
\item Sort $L$ and take the first $k$ samples with smallest distances
\item Let $y_l$ be the majority class of the $k$ samples
\end{enumerate}
\caption*{Figure 3: Pseudocode of Inferencing with KNN}
\end{figure}

\section*{Experimental Setup}

\subsection*{Protein Data Bank}
The Protein Data Bank (PDB) is a freely accessible, interoperating archive of structured data and metadata of biological macromolecules (proteins and nucleic acids) for research and education across the sciences funded by the National Science Foundation, the National Institutes of Health, and the Department of Energy. The experimental data, obtained by biologists using methods such as X-ray crystallography, NMR spectroscopy, and cryo-electron microscopy, are open-source on the Internet through its member organizations. The PDB is set as the gold standard for many areas of biology, including structural genomics, and other databases use protein structures from the PDB. In fact, many scientific journals and funding agencies require researchers to submit their data to the PDB. Hence, the database is rich with thousands of samples of proteins, each with features that may be useful as covariates for classification. These features include the primary structure of each protein, which will be of interest when building our sequence-based deep learning models. 

\begin{figure}[H]
    \centering
    \includegraphics[scale = 0.4]{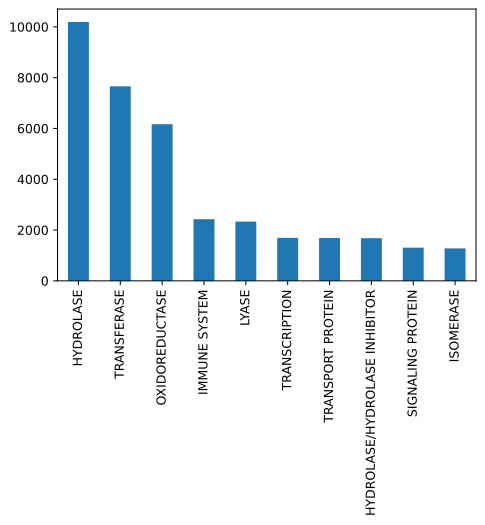}
    \caption{Figure 3: Top Frequencies of Protein Families in Protein Data Bank}
    \label{fig:my_label}
\end{figure}

The raw PDB dataset includes 60757 non-unique samples of proteins, DNA, and RNA with 16 total features including publication data and sequencing information. After removing redundant and non-protein samples and reducing the features to only include experimental and primary protein structure data, proteins with NaN data points or that were not a part of the ten most frequent protein families were also excluded from the analyses. Of the remaining 36,411 proteins, 7,282 are left for testing, while the remaining 29,129 samples are used for training and validation with a 15\% split. The data pipeline for this work is of the form:

\begin{equation}
    f(X) = f_{\text{model}}(f_{\text{onehot}}(f_{\text{tokenization}}(f_{\text{kmer}}(X))))
\end{equation}

\subsection*{Deep Learning Models}
\subsubsection*{RNN}
In this work, the RNN implemented is a bidirectional LSTM architecture using the Tensorflow framework. This architecture is well-studied and academic literature on the topic is broad, so we will instead focus on how we transform the protein sequencing data into a format usable for these types of models. \\
Recall that RNN models require a sequence of tokens, where each token is composed of letters of an alphabet, as input. The current data does not satisfy this condition as long sequences of proteins. Hence, we can attempt to create a language for our model by breaking up a protein sequence into tokens using the following trick. First, let us define the alphabet of our language to be the set of amino acids that comprise a protein; the length of our alphabet is thus twenty. Next, we define tokens as equal-length k-mers that are generated by sliding a window of size $k$ across the sequence, where each step $s$ constitutes a new word in the sentence. Tada! Our protein sequence has transformed into a sentence of amino-acid words that hold intrinsic information about the protein in both the amino acids and the ordering of the tokens.
With a tokenized sequence, we now move onto embedding. The motivation behind this step is that our language model requires a matrix-representation of the inherent meaning of the tokens that can be processed by the hidden cells of the RNN. Essentially, we would like to project the information of a string token into euclidean space. To do this, we initialize each token as a random $1xd$ vector, where $d$ is the number of dimensions of the euclidean space, and train a feed forward neural network to learn the true value of these vectors. We can reformat our protein sequence into the form $[(a,c), b]$, where $b$ represents the current token, $a$ represents the $t$ tokens before it, and $b$ represents the $t$ tokens after it. We call $(a,c)$ the context vectors around $b$ for which the "lemma" neural network model will attempt to learn the vector representation of. Formally, this form of learning is known as word2vec, as we find a mapping of words to vectors, and this particular context-based algorithm is known as the skip-gram architecture \cite{bib29}.\\
Our character-based language model operates on protein sequence data tokenized on the level of amino acids. Each protein sequence can now be represented as an $n x d$ input matrix, where each $n$ represents the length of the sequence and $d$ represents the dimension of the embedding space. The architecture of this model is of the form:

\begin{equation}
    f(X) = f_{\text{mlp}}(f_{\text{BDLSTM}}(f_{\text{embedding}}(X)))
\end{equation}

\subsubsection*{CNN}
In this work, we implement a CNN for protein family classification using the Tensorflow framework. Our model is composed of five stages: convolution, rectification, pooling, dropout, and a multi-layer perceptron. Each convolution stage consists of 128 filters of varying sizes between four and six such that each filter learns a different pattern of the data. The size and number of filters is a hyperparameter at the discretion of the authors, and we found that these particular sizes and number of filters worked well for this task. The rectification stage pairs with the prior stage to apply a non-linearity function called rectified linear unit (ReLU), which is an elementwise operation that sets all negative values to approach zero. Here, we utilize leaky ReLU instead of the traditional ReLU to avoid vanishing gradient issues that may persist with the size of our model. 

\begin{equation}
f_{\text{leaky ReLU}} (z) = 
    \begin{cases}
      \alpha x & \text{if $x \leq 0$}\\
      x & \text{if $x>0$}
    \end{cases}       
\end{equation}

Next, we use temporal maxpooling, which selects the maximum values in a certain window, to learn translational invariant features and denoise the output of prior stages. For this CNN model, we set a window size of 2, which conservatively allows for both denoising of the output of prior stages while also allowing the model to have enough data to make an informed prediction. Following pooling is dropout \cite{bib28}, which helps to prevent overfitting by randomly zeroes the inputs to the next layer during training with a set probability of 0.2. It should be noted that this method closely resembles ensemble techniques in classical machine learning like bagging or model averaging which are heavily utilized in bioinformatics. Finally, this output is fed into a feedforward neural network to learn a classification function mapping to protein family labels. Each layer in the MLP maps its input to a hidden feature space, known as hidden layers, and the final layer learns to map from the hidden space of the last hidden layer to the class label space through a softmax function. This model is trained using the Adam optimizer to minimize our categorical cross-entropy loss function. The architecture of this model is of the form:

\begin{equation}
f(X) = f_{\text{mlp}}(f_{\text{dropout}}(f_{\text{pool}}(f_{\text{relu}}(f_{\text{conv}}(X)))))
\end{equation}

\subsection*{Baseline Models}
The baseline models utilize the experimental data available, as opposed to the sequencing data, as the experimental data matches the input parameters for the kNN classifier and the multinomial regression model. As such, the 36,411 protein samples are feature-normalized to a standard normal distribution. Both models are implemented in Python. The kNN classifier was run to produce an elbow graph for the best value of $k$. In Figure 2, we see that the optimal value of $k$ is 3. No hyperparameter tuning is needed for multinomial regression.

\subsubsection*{Feature Selection}
Extremely Randomized Trees Classifiers (Extra-Treess Classifier) are an ensemble learning technique that aggregates the results of multiple collected de-correlated decision trees to find features that are most important in the data set, a technique first utilized by ~\cite{bib11}. During the construction of the forest, the normalized total reduction of feature split decision for each feature is computed, also called Gini importance of the feature, describes how "important" the particular feature is to the outcome variable. Our results from running the Extra-Trees Classifier on our experimental data features are presented in Figure 5. We found that the Gini index values for each feature is approximately 0.11, except for the residue count and molecular weight features which have Gini indices values closer to 0.20, we utilize all features for the classical machine learning models.\\

\begin{figure}[H]
\center
\includegraphics[scale = 0.35]{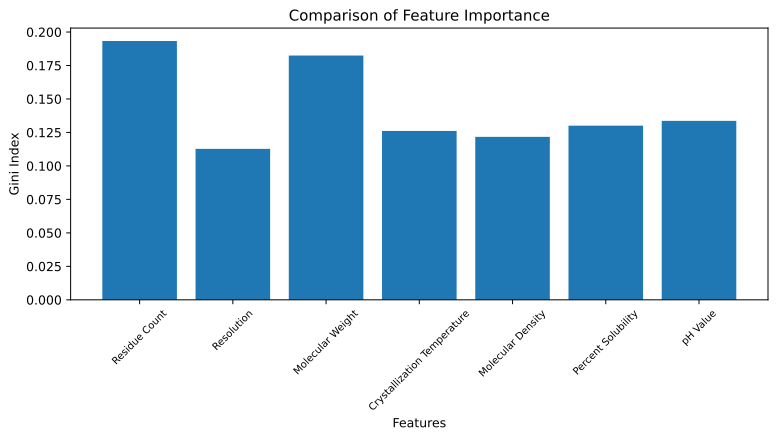}
\caption*{Figure 5: Gini Indices for Feature Selection}
\end{figure}

\section*{Results}

Three metrics were utilized for evaluating the performance of the classifiers on the data, each with their own scoring schemes: accuracy, precision, and recall. Accuracy measures the ratio of correct classifications (true positives and true negatives) to the total number of samples. In other words, accuracy = $\frac{TP + TN}{TP + TN + FP + FN}$. Precision is the proportion of the classifications of proteins to a particular protein family that actually belong to that family. In other words, precision = $\frac{TP}{TP + FP}$. Recall is the proportion of proteins in a protein family that were classified into the right family. In other words, recall = $\frac{TP}{TP + FN}$. Because these evaluation metrics are used on multi-class classification models, accuracy, precision, and recall are calculated for each class and averaged to get the final metrics of that particular model.

\begin{figure}[H]
\center
\begin{table}[H]
\center
\begin{tabular}{|l|l|l|l|}
\hline
                             & \textbf{Accuracy} & \textbf{Precision} & \textbf{Recall} \\ \hline
\textbf{Multinomial Regression}  & 0.29             & 0.23              & 0.29           \\ \hline
\textbf{K-Nearest Neighbors} & 0.40             & 0.41              & 0.40           \\ \hline
\textbf{Bi-Directional LSTM} & 0.79             & 0.75              & 0.77           \\ \hline
\textbf{CNN} & \textbf{0.83}             & \textbf{0.79}              & \textbf{0.81}           \\ \hline
\end{tabular}
\end{table}

\caption{Figure 6: Performance Metrics of Classifiers}
\end{figure}

 Due to the contrast in results between the deep learning models and classical ML models, we begin our discussion of the results of the two separately.
 Of the ML models, the KNN classifier had the higher accuracy of 0.40, which indicates that proteins within the same family are closer in Euclidean space than proteins in other families. Looking back at our data, this result makes sense because the number of samples in our data is high, which means that there are more proteins in the same family near each other, resulting in a higher accuracy. The KNN classifier also had the higher precision of 0.41. This shows that the data's relatively lower dimensionality also aided the performance of the KNN, for if the dimensionality of the data were high, the model would be more susceptible to adding $x_j$'s of a different protein family into $x_i$'s $k$ nearest neighbors since the curse of dimensionality makes it more difficult for the model to differentiate between different points in space using distance~\cite{bib7}. Not surprisingly, the recall of the KNN classifier was also the highest. Again, the sheer number of samples played a large role in this; since there are more proteins of the same family neighboring each other, a higher proportion of the proteins within the same family will be classified correctly, thus leading to a higher precision.\\
 The multinomial regression model was unable to perform as well as the KNN classifier. The first reason for this is the class imbalance of the data. Of the 36411 proteins, 10190 were from a single protein family, hydrolase, which is more than the frequencies of the bottom six proteins \textit{combined}. Thus, the model learned to guess that any protein it saw most likely came from the hydrolase family. In fact, this is clear from the model's accuracy of 0.29, which is just slightly above the relative hydrolase frequency in the data of 0.28. Indeed, this also explains the precision score of 0.23; the overall model precision is an average of the precision of each protein family, most of which will be lower than $0.29$ and thus lower the average. The precision for hydrolase is $0.30$, which means that the model is guessing for a majority of the proteins that they belong to the hydrolase protein family, for if this weren't the case, then the hydrolase precision would be much higher. This is supported by the near-zero precision scores of the other 9 protein families.\\
 
 \begin{figure}[H]
    \centering
    \subfloat[\centering CNN Confusion Matrix]{{\includegraphics[width=7cm]{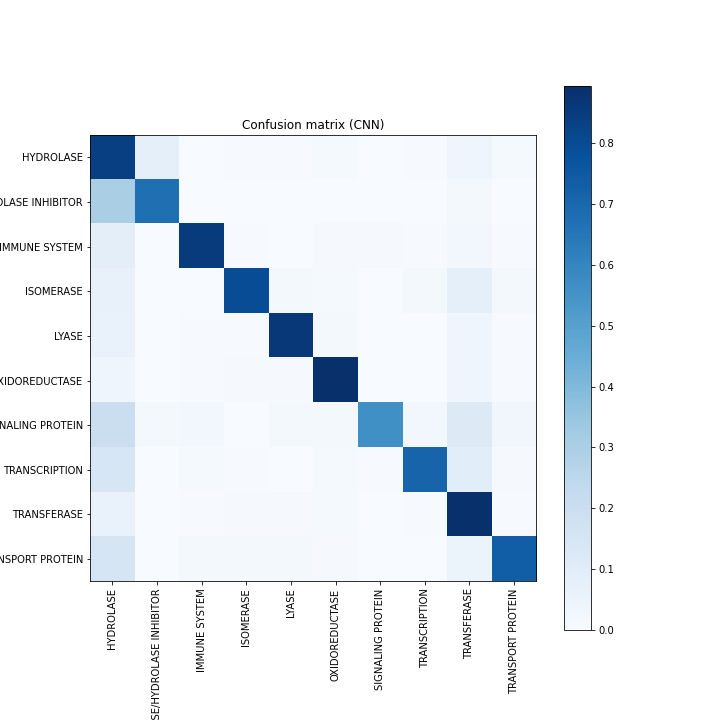} }}%
    \qquad
    \subfloat[\centering LSTM Confusion Matrix]{{\includegraphics[width=7cm]{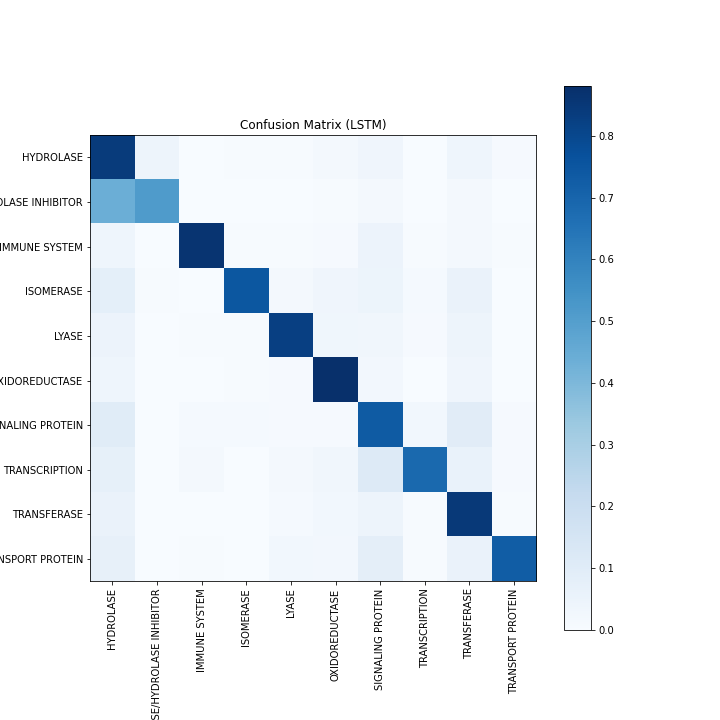} }}%
    \caption{Figure 7: Deep learning models confusion matrices}%
    \label{fig:example}%
\end{figure}
 
 The deep learning models presented in this work perform exceptionally better than both of the classical machine learning approaches. The CNN performed the best with an accuracy of 0.83, while the LSTM lagged only slightly with an accuracy of 0.79. In fact, the evaluation metric of each deep learning model is greater than the two classical machine learning models' evaluation metrics combined. There are several reasons for this discrepancy. The deep learning models are severely more complex than the classical machine learning models, and because protein family classification is such a difficult task, it is intuitive that classical machine learning models do not have the complexity bandwidth to learn all of the intrinsic patterns within the data. In comparison to the classical machine learning models, which only have a handful of trainable parameters (kNN does not have even have a training phase, whereas multinomial regression has only $f \cdot k = 70$ trainable parameters), the convolutional neural network has almost 4 million trainable parameters and the bi-directional LSTM model has over 41 million trainable parameters. The added complexities of the deep learning models allow for ordered representation of the data, whether spatial or sequential, that is not seen in the classical machine learning models. Furthermore, it should be noted that the models use different data as input. While the classical ML models used experimental data, the deep learning models used protein sequence data; this matters because the protein sequence data has an inherent order to it, unlike the experimental data, that conveys information.

\begin{figure}[H]
    \centering
    \subfloat[\centering CNN Accuracy]{{\includegraphics[width=7cm]{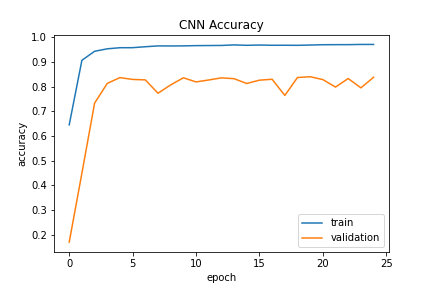} }}%
    \qquad
    \subfloat[\centering LSTM Accuracy]{{\includegraphics[width=7cm]{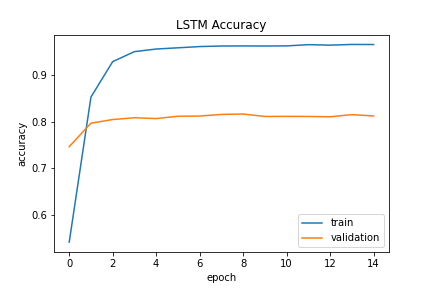} }}%
    \caption{Figure 8: Deep learning models training and validation accuracy per epoch}%
    \label{fig:example}%
\end{figure}

\section*{Discussion}
In this work, we have presented a comparison of several deep learning and machine learning models, including a recurrent neural network, a convolutional neural network, multinomial regression, and k-nearest neighbor, for protein classification on sequential and experimental data from the Protein Data Bank. Our findings show that despite multinomial regression fails to provide an edge over the k-nearest neighbors classifier. There are several reasons why this may be the case, including class imbalance, low feature space, and high number of samples. Hyperparameter tuning of the multinomial regression model, such as trying different optimizers, did not make any significant change in the performance of the model. Another observation is that the data points collected may not have enough significant figures to differentiate the samples well enough. For example, the observations within the crystallization temperature feature (in Kelvin) only contain one decimal place, which certainly is not enough to differentiate 36000 proteins into 10 classes, especially since deviation outside of the optimal temperature leads to protein denaturation. This aspect of the data will certainly improve the performance of the KNN since neighbors of a sample may even have the same measurements, but will decrease the performance of the multinomial regression model since a high number of the same sample measurements leads to guessing the majority class. This is supported by the plot of the effect of $k$ on KNN accuracy, which shows that as $k$ increases, the accuracy of the classifier decreases; in fact, $k=1$ has nearly the same accuracy as the top accuracy, which shows that looking at the nearest neighbor is equally a good choice for determining classifications. A biological underpinning for our results may also be that the protein families we attempted to classify the samples into may be too similar. These proteins were chosen since they were the most frequent in the Protein Data Bank, so it is feasible that there exists underlying relations between the families that skew the performance of the models. For example, if some of the 10 protein families are themselves evolutionarily or functionally related, then it makes sense that the multinomial regression model outputs guesses since it is impossible to distinguish between classes.
Comparing the two deep learning models is difficult. This is because there is no intuitive reasons for the difference in results between the two; that is, what gains are made in inference performance are lost in interpretability of the results. What can be said, however, is that it may be possible that both models overfit the protein sequence data. This is suggested by the near immediate convergence of the models to their best validation accuracy scores within the first five epochs of around 0.80. Yet, during training, the CNN averaged an accuracy of 0.95 while the LSTM averaged an accuracy of 0.94. This may indicate that the models overfit the protein sequence data with their millions of parameters and thus is why we see this immediate convergence. To combat this, future work should be done on regularization of RNNs and CNNs on bioinformatics data, as well as on the interpretability of these deep learning models such that the inference procedure of these "black-box" models can be better understood: we know not the underlying biological mechanisms within the sequence that allow for these classifications, which make inference from deep learning models, however accurate, not useful. We hypothesize that the CNN models perform better than the RNN models on bioinformatics data because of the varying degree and flexibility of the kernels, in their size, stride, padding, and pooling techniques. From this flexibility, CNNs are able to learn short-term and long-term dependencies across the sequence that RNN models generally struggle with, including LSTMs. The introduction of transformers in recent years seem like an interesting improvement to RNNs that may level the playing field. For the machine learning models, we would like to examine the effects of data synthesis and random over-sampling of the minority classes. We hypothesize that this will greatly boost the performance of the model since a class-balanced dataset may prevent guessing of the majority class. We would also like to better understand how the data is collected, what machinery is used, and how precise and specific these measurements are. \\
In summary, we have presented an analysis of several deep learning and classical machine learning models for protein classification on Protein Data Bank Data and shown that deep learning models, in particular convolutional neural networks, are superior over classical machine learning techniques due to their added complexity and ability to understand sequential information on PDB data. Despite the power of these models, constraints of interpretability leave much to be desired in the way of meaningful classifications, a possible direction for future work.

\section*{Acknowledgements}
This work was done as part of graduate coursework for CSCI 2820: Algorithmic Foundations of Computational Biology, under the instruction of distinguished Professor \textit{Sorin Istrail}. I would like to thank him for an exceptional semester and appreciate the feedback from fellow students during the project’s development. I look forward to learning more under his guidance in the future.

\bibliographystyle{unsrtnat}


\begin{thebibliography}{1}
  \bibitem{bib1}
  Dongardive, Jyotshna, and Siby Abraham. “Protein Sequence Classification Based on N-Gram and K-Nearest Neighbor Algorithm.” Computational Intelligence in Data Mining—Volume 2, edited by Himansu Sekhar Behera and Durga Prasad Mohapatra, vol. 411, Springer India, 2016, pp. 163–71, doi:10.1007/978-81-322-2731-1\_15.
  
  \bibitem{bib2}
  Strodthoff, Nils, et al. “UDSMProt: Universal Deep Sequence Models for Protein Classification.” Bioinformatics, edited by Yann Ponty, vol. 36, no. 8, Apr. 2020, pp. 2401–09, doi:10.1093/bioinformatics/btaa003.
  
  \bibitem{bib3}
  Czepiel, Scott A. Maximum Likelihood Estimation of Logistic Regression Models: Theory and Implementation. p. 23.
  
  \bibitem{bib4}
  Böhning, Dankmar. “Multinomial Logistic Regression Algorithm.” Annals of the Institute of Statistical Mathematics, vol. 44, no. 1, Mar. 1992, pp. 197–200, doi:10.1007/BF00048682.
  
  \bibitem{bib5}
  Royer, Clément W., et al. “A Newton-CG Algorithm with Complexity Guarantees for Smooth Unconstrained Optimization.” Mathematical Programming, vol. 180, no. 1, Mar. 2020, pp. 451–88, doi:10.1007/s10107-019-01362-7.
  
  \bibitem{bib6}
  Murphy, K. P. Machine Learning: A Probabilistic Perspective. MIT Press, 2012, https://books.google.com/books?id=RC43AgAAQBAJ.
  
  \bibitem{bib7}
  Jiang, Shengyi \& Pang, Guansong \& Wu, Meiling \& Kuang, Limin. (2012). An Improved k-Nearest Neighbor Algorithm for Text Categorization. Expert Systems with Applications. 39. 1503-1509. 10.1016/j.eswa.2011.08.040.
  
  \bibitem{bib8}
  R. Duncan Luce. Individual Choice Behavior: A Theoretical Analysis. John Wiley \& Sons, 1959.
  
  \bibitem{bib9}
  Jurafsky, Daniel \& Martin, James. (2008). Speech and Language Processing: An Introduction to Natural Language Processing, Computational Linguistics, and Speech Recognition. 
  
  \bibitem{bib10}
  Abu Alfeilat, Haneen Arafat, et al. “Effects of Distance Measure Choice on K-Nearest Neighbor Classifier Performance: A Review.” Big Data, vol. 7, no. 4, Dec. 2019, pp. 221–48, doi:10.1089/big.2018.0175.
  
  \bibitem{bib11}
  Geurts, Pierre, et al. “Extremely Randomized Trees.” Machine Learning, vol. 63, no. 1, Apr. 2006, pp. 3–42, doi:10.1007/s10994-006-6226-1.
  
  \bibitem{bib12}
  Jha, Anupama, Matthew R. Gazzara, and Yoseph Barash. "Integrative deep models for alternative splicing." Bioinformatics 33.14 (2017): i274-i282.
  
  \bibitem{bib13}
  Quang, Daniel, Yifei Chen, and Xiaohui Xie. "DANN: a deep learning approach for annotating the pathogenicity of genetic variants." Bioinformatics 31.5 (2015): 761-763.
  
  \bibitem{bib14}
  Liu, Feng, et al. "PEDLA: predicting enhancers with a deep learning-based algorithmic framework." Scientific reports 6.1 (2016): 1-14.
  
\bibitem{bib15}
  Li, Yifeng, Wenqiang Shi, and Wyeth W. Wasserman. "Genome-wide prediction of cis-regulatory regions using supervised deep learning methods." BMC bioinformatics 19.1 (2018): 1-14.
  
\bibitem{bib16}
  Angermueller, Christof, et al. "DeepCpG: accurate prediction of single-cell DNA methylation states using deep learning." Genome biology 18.1 (2017): 1-13.
  
\bibitem{bib17}
  Quang, Daniel, and Xiaohui Xie. "DanQ: a hybrid convolutional and recurrent deep neural network for quantifying the function of DNA sequences." Nucleic acids research 44.11 (2016): e107-e107.
  
\bibitem{bib18}
Degroeve, Sven, et al. "Feature subset selection for splice site prediction." Bioinformatics 18.suppl\_2 (2002): S75-S83.

\bibitem{bib19}
Bucher, P. “Weight matrix descriptions of four eukaryotic RNA polymerase II promoter elements derived from 502 unrelated promoter sequences.” Journal of molecular biology vol. 212,4 (1990): 563-78. doi:10.1016/0022-2836(90)90223-9

\bibitem{bib20}
Heintzman, N., Stuart, R., Hon, G. et al. Distinct and predictive chromatin signatures of transcriptional promoters and enhancers in the human genome. Nat Genet 39, 311–318 (2007).

\bibitem{bib21}
Segal, Eran et al. “A genomic code for nucleosome positioning.” Nature vol. 442,7104 (2006): 772-8. doi:10.1038/nature04979

\bibitem{bib22}
Ohler, U., Liao, Gc., Niemann, H. et al. Computational analysis of core promoters in the Drosophila genome. Genome Biol 3, research0087.1 (2002)

\bibitem{bib23}
Alipanahi, B., Delong, A., Weirauch, M. et al. Predicting the sequence specificities of DNA- and RNA-binding proteins by deep learning. Nat Biotechnol 33, 831–838 (2015). 

\bibitem{bib24}
Kelley, David R., Jasper Snoek, and John L. Rinn. "Basset: learning the regulatory code of the accessible genome with deep convolutional neural networks." Genome research 26.7 (2016): 990-999.

\bibitem{bib25}
Zhou, J., Troyanskaya, O. Predicting effects of noncoding variants with deep learning–based sequence model. Nat Methods 12, 931–934 (2015).

\bibitem{bib26}
Ritambhara Singh, Jack Lanchantin, Gabriel Robins, Yanjun Qi, DeepChrome: deep-learning for predicting gene expression from histone modifications, Bioinformatics, Volume 32, Issue 17, 1 September 2016.

\bibitem{bib27}
Qi, Yanjun et al. “A unified multitask architecture for predicting local protein properties.” PloS one vol. 7,3 (2012): e32235. doi:10.1371/journal.pone.0032235

\bibitem{bib28}
Srivastava, Nitish \& Hinton, Geoffrey \& Krizhevsky, Alex \& Sutskever, Ilya \& Salakhutdinov, Ruslan. (2014). Dropout: A Simple Way to Prevent Neural Networks from Overfitting. Journal of Machine Learning Research. 15. 1929-1958. 
  
 \bibitem{bib29}
 Mikolov, T., Sutskever, I., Chen, K., Corrado, G. S., \& Dean, J. (2013). Distributed Representations of Words and Phrases and their Compositionality. In C. J. Burges, L. Bottou, M. Welling, Z. Ghahramani, \& K. Q. Weinberger (Eds.), Advances in Neural Information Processing Systems (Vol. 26). Curran Associates, Inc. 
  
  




















  

\end{thebibliography}

\section*{Supplementary Data}

\begin{figure}[H]
    \centering
    \includegraphics[scale = 0.5]{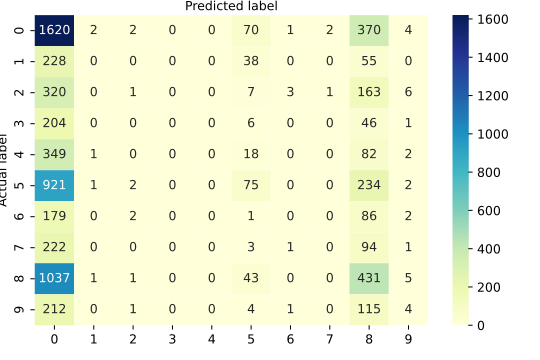}
    \caption{Figure 9: Multinomial Regression Confusion Matrix}
    \label{fig:my_label}
\end{figure}

\end{document}